\documentclass{article}
\usepackage{spconf,amsmath,graphicx,hyperref}

\usepackage{cite}
\usepackage{amssymb,amsfonts}
\usepackage{algorithmic}
\usepackage{textcomp}
\usepackage{xcolor}

\usepackage{cleveref}
\usepackage{url}
\usepackage[table]{xcolor}
\usepackage{booktabs}
\usepackage{multirow}
\usepackage[most]{tcolorbox}
\usepackage{todonotes}

\newtcblisting[blend into=figures]{promptbox}[1][]{
  arc=2mm,
  colback=gray!6!white,
  colframe=gray!75!black,
  fonttitle=\bfseries\sffamily\small,
  listing only,
  listing options={
    basicstyle=\ttfamily\small,
    breaklines=true,
    columns=fullflexible
  },
  #1
}

\crefname{appendix}{\S}{\S}
\Crefname{appendix}{\S}{\S}
\crefname{section}{\S}{\S}
\Crefname{section}{\S}{\S}
\crefname{subsection}{\S}{\S}
\Crefname{subsection}{\S}{\S}

\def\BibTeX{{\rm B\kern-.05em{\sc i\kern-.025em b}\kern-.08em
    T\kern-.1667em\lower.7ex\hbox{E}\kern-.125emX}}

\let\oldthebibliography\thebibliography
\let\endoldthebibliography\endthebibliography
\renewenvironment{thebibliography}[1]{%
  \oldthebibliography{#1}%
  \setlength{\itemsep}{0pt plus 1pt}%
  \setlength{\parskip}{1.5pt}%
}{%
  \endoldthebibliography%
}

\title{\textsc{Logbook}: Extremely Long-form Audio Event Understanding}

\name{
\begin{tabular}{c}
    \itshape Kwanghee Choi$^{\bullet,*}$, Suwon Shon$^*$, Dmitriy Serdyuk$^*$, Guitang Lan$^*$, Chao-Wei Huang$^*$, \\
    \itshape Mohammad Sadegh Rasooli$^*$, Sangeeta Srivastava$^*$, Zhaojiang Lin$^*$, Saurabh Adya$^*$, Ming Sun$^*$
  \end{tabular}
}
\address{$^{\bullet}$UT Austin, USA \quad $^*$Meta Reality Labs, USA}

\begin{document}
\ninept
\maketitle

\begin{abstract}
Audio benchmarks are built around short, pre-segmented clips, limiting model design to brief inputs or fixed vocabularies.
To close this gap, we introduce \textsc{Logbook}, a benchmark for hour-scale audio understanding, with recordings ranging from ten minutes to six days.
Given a continuous audio recording and an event label vocabulary, a system must predict a gap-free segmentation with an event label and a description per segment.
We compare 52 systems, end-to-end and cascaded, and ablate fine-tuning, context length, and reasoning budget.
We find the task tractable, though the best systems remain below the human reference.
Also, over-segmentation is pervasive, and fine-tuning partially mitigates it.
Finally, end-to-end are often better than cascaded systems, but degrades with longer context.\footnote{Supplementary materials, including \Cref{apdx}, are available at \url{https://github.com/facebookresearch/logbook}.}\end{abstract}

\begin{keywords}
long audio, audio event segmentation, egocentric audio, audio-language model, benchmark.
\end{keywords}

\section{Introduction}
Hearing is continuous.
Throughout the day, we experience an unbroken acoustic stream and parse into segments such as conversation, meal, or commute, where wearable devices are now able to capture \cite{lin2026wearvox,paruchuri2025egotrigger}.
However, existing audio event datasets often consist of short clips, already cut to contain a single event \cite{gemmeke2017audio,serizel2020sound,DrossosLV20}.
Some egocentric datasets provide continuous records \cite{grauman2024ego4d,yang2025egolife}, but none is designed for long audio: many often run under an hour \cite{kitani2011fast,aghazadeh2011novelty,youhome2022,fredianelli2025environmental}, cover only a specific activity such as kitchen work \cite{damen2020epic,perrett2025hd}, or treat audio as secondary to video \cite{grauman2024ego4d,yang2025egolife} (see \cite{yang2025egolife} for a survey).
Finally, long-audio benchmarks reach for duration, yet skew toward speech \cite{ahia2025blab}, and capped at 20 and 30 minutes \cite{luo2026chronosaudio,shao2026listening}.

The gap influences model design.
Sound event detectors are often built and evaluated on short recordings with a fixed vocabulary \cite{mesaros2021sound}.
Audio captioning models commonly emit a single sentence regardless of input length \cite{mei2022automated}.
Even modern audio language models with longer context are evaluated on at most five minutes \cite{ghosh2026audio,xu2025qwen3}.

We argue that the task should mimic human listening: understanding continuous audio as a segmented record, leading to the \textsc{Logbook} benchmark.
\textbf{Task.} We newly define long-form audio understanding as a gap-free segmentation in which every segment carries an event label and a description (\Cref{fig:prompt}).
\textbf{Benchmark.} We curate publicly available datasets into a unified format, with uninterrupted recordings from ten minutes to six days (\Cref{tbl:dataset}), and design an evaluation suite scoring segmentation and description quality, with a human baseline.
\textbf{Extensive comparisons.} We compare 48 cascaded systems pairing six audio captioning models with eight text LLMs, and four end-to-end (E2E) audio language models. We also ablate fine-tuning, context window size, and reasoning budget.

\begin{promptbox}[title=Example input and output schema, boxsep=1pt, top=2pt, bottom=2pt, label={fig:prompt}]
Input = {
  "audio": ...,  // Raw audio
  "label": {  // Class definitions
    "leisure": "Enjoyable activities.",
    "traveling": "Commuting or travel between activities.", ... }
}
Output = [{
  "start": 0, "end": 24,  // In minutes
  "event": "traveling",
  "description": "The wearer is commuting by taking the subway."
}, ... // Other actions]
\end{promptbox}

We conclude that gap-free segmentation is tractable from audio alone.
However, systems often over-segment, while fine-tuning partially mitigates such.
Also, E2E are often stronger than cascaded systems in both performance and compute, yet degrade with longer windows. 
Reasoning yields limited gains, mainly by improving instruction following rather than audio understanding.

\section{The Logbook Benchmark}
\vspace{-0.2em}\subsection{Task definition (\Cref{fig:prompt})}\vspace{-0.2em}
\noindent\textbf{Input.} A continuous recording, ranging from minutes to days, and a prompt specifying the output format and a fixed vocabulary of event labels with their definitions (full prompt in \Cref{apdx:prompts}).

\noindent\textbf{Output.} Non-overlapping, contiguous segments that fully cover the recording, each with one event label and a free-text description.

\noindent\textbf{Rationale.} The label vocabulary sets the granularity of segmentation.
For example, cooking and eating can collapse into a single \textit{food} segment.
The description carries the information that the label omits, so that the pair accounts for the full content of the segment.
Requiring full coverage also keeps systems comparable: every second is assigned, so a system cannot raise precision by abstaining.

\vspace{-0.5em}\subsection{Datasets}
\noindent\textbf{Ego4D} \cite{grauman2024ego4d} is an egocentric video dataset covering daily-life activity across a wide range of scenarios.
We use their Summary and Moments tiers.
In the Summary tier, two human annotators independently cover each recording, writing one to three sentences per five-minute window.
The Moments tier localizes individual actions from a 110-class vocabulary, where most last under a minute.

\noindent\textbf{EgoLife} \cite{yang2025egolife} records six participants living together for one week.
We use its dense captions: first-person Mandarin, averaging 2.7s per caption.
Sessions are formed by cutting audio at gaps over one second and merging caption fragments less than a minute apart.

\noindent\textbf{SINS} \cite{dekkers2017sins} captures six days of a single inhabitant, recorded by twelve microphone arrays throughout a house.
It provides contiguous event labels over 17 event classes, but no descriptions.

\vspace{-0.5em}\subsection{Dataset preprocessing}
\begin{table}[t]
\centering
\caption{Dataset statistics. \# sess. is the number of uninterrupted sessions and \# subj. is the number of unique recorded subjects.}
\label{tbl:dataset}
\resizebox{\linewidth}{!}{
\begin{tabular}{llllllll}
\toprule
\textbf{Source} & \textbf{Split} & \textbf{\# sess.} & \textbf{Avg. len.} & \textbf{Tot. len.} & \textbf{Label} & \textbf{Domain} & \textbf{\# subj.} \\
\midrule
\textbf{Ego4D}  & Train & 1350 & 24.7 min & 555.8 h & Inferred & Egocentric & 250 \\
\textbf{Ego4D}  & Val   & 247  & 26.1 min & 107.6 h & Inferred & Egocentric & 71  \\
\textbf{Ego4D}  & Test  & 163  & 30.2 min & 82.1 h  & Inferred & Egocentric & 55  \\
\midrule
\textbf{EgoLife}& Test  & 170  & 84.9 min & 240.6 h & Inferred & Egocentric & 6   \\
\midrule
\textbf{SINS}   & Test  & 1    & 148.9 h  & 148.9 h & Provided & Smart home & 1   \\
\bottomrule
\end{tabular}
}
\vspace{-1.5em}
\end{table}
\noindent\textbf{Ego4D.}
We use OLMo-2-1124-7B-SFT \cite{walsh20252} for all text preprocessing.
Each five-minute narration slice passes three stages.
\textit{Clean} stage formats the raw narration into a third person summary.
\textit{Atomic fact generation (AFG)} stage decomposes the cleaned summary into atomic facts, following \cite{min2023factscore,lage2025openfactscore}.
\textit{Classify} stage assigns one of six activity classes (productive, food, leisure, purchasing, traveling, other), adapted from the American Time User Survey \cite{bls2026atus} (full definition in \Cref{apdx:prompts}).
Then, adjacent slices with the same class are merged.
Ego4D splits are assigned 60/20/20 by hashing the participant identifier, so the assignment is deterministic and no participant appears in two splits.
Ego4D's recordings are unevenly distributed across capture sites, and we keep only the four sites with longer than ten test set hours: `{iiith}', `{minnesota}', `{kaust}', and `{frl\_track\_1\_public}'.
Finally, for the test split, we filter the recordings that are shorter than ten minutes or have fewer than two events.

\noindent\textbf{EgoLife.}
We follow the similar pipeline to Ego4D.
We first translate the dense captions from Mandarin with NLLB-200-distilled-600M \cite{costa2022no}, then group them into five-minute windows to match Ego4D's granularity.
The label vocabulary for \textit{Classify} stage is reused from Ego4D unchanged, and each caption is directly used as atomic facts.
Finally, we apply the same test set filter to Ego4D.

\noindent\textbf{SINS.}
We create a mono-stream audio using the ground-truth annotation.
We select the array nearest the inhabitant by crossfading over two seconds.
Unannotated intervals are filled with absence (17\% of full audio), making the label timeline fully contiguous.

\subsection{Evaluation}
\noindent\textbf{Segmentation.}
\textit{Boundary F1 (bF1)} measures whether boundaries are placed correctly.
A boundary counts as matched if it falls within 150s.
Harmonic mean is calculated between precision and recall.
\textit{Event error rate (evER)} discards length and compares only the predicted label sequence to the reference by the Levenshtein distance, being a diagnostic for over-segmentation.
\textit{Frame-level Accuracy (fAcc)} reports the fraction of correctly labelled frames, penalizing boundary and label errors jointly.
We use it as the headline metric.
bF1 and fAcc are micro-averaged, and evER is macro-averaged.

\noindent\textbf{Description.}
Similar to FactScore \cite{min2023factscore,lage2025openfactscore}, predicted descriptions are decomposed into atomic facts by the same OLMo model, and each fact is compared against the reference facts with BART fine-tuned on text entailment (BART-large-MNLI \cite{lewis2020bart}).
Predicted and reference fact pairs within the temporally overlapping segments are compared, and a pair counts as matched when the entailment probability exceeds 0.5.
\textit{Recall (dR)} is the fraction of reference facts entailed by some predicted fact, and \textit{Precision (dP)} the fraction of predicted facts entailed by some reference fact, where both are micro-averaged.
\textit{F1 (dF1)} is their harmonic mean, the headline metric.

\noindent\textbf{Compute.}
We report \textit{peta-FLOPs (PFLOPs)} to process the first 24 h of SINS.
We first measure PFLOPs of one chunk, multiply by the number of chunks, then add an autoregressive approximation of \cite{kaplan2020scaling} (2 $\times$ \# of generated tokens $\times$ \# of active parameters).

\noindent\textbf{Downstream utility.}
We additionally measure two metrics for Ego4D.
\textit{Hit rate} is recall@1 against the Moments tier: for each annotated moment, whether any temporally overlapping predicted fact entails it.
\textit{Redundancy (rddc)} is the fraction of predicted facts entailed by another fact within the same segment.

\noindent\textbf{Human reference.}\label{ss:human}
Ego4D has two annotation passes, so we score each against the other and pool both directions.
Note that this is a proxy: the labels were inferred by an LLM and carry its errors; both passes sit on a five-minute grid, earning no partial credit for bF1 under 150s tolerance, unlike minute-resolution systems; annotators vary in detail; and annotators saw video, which systems do not.

\vspace{-0.5em}\section{Systems}
We compare two system designs.
End-to-end (E2E) systems consume the audio directly and predict the segments, each with an event label and a description.
Cascaded (Casc.) systems first caption short audio windows, then pass the captions to a text LLM to emit the same output format.
Detailed prompts are in \Cref{apdx:prompts}.

\vspace{-0.5em}\subsection{End-to-end systems}
We use four multimodal models with long context windows for audio.
\textit{Qwen 2.5-Omni-7B} \cite{xu2025qwen25} (Qwen 2.5-o) reads and generates both text and speech, with strong general audio understanding.
\textit{Qwen 3-Omni-30B-A3B-Instruct} \cite{xu2025qwen3} (Qwen 3-o) is its successor, a mixture-of-experts model \cite{ShazeerMMDLHD17} with 30B total and 3B active parameters.
\textit{Audio Flamingo 3} \cite{ghosh2026audio} (AF 3) targets audio and text only, aligning a pretrained Whisper encoder \cite{radford2023robust} with a Qwen 2.5-7B decoder \cite{yang2025qwen2}.
\textit{Gemini 3.5 Flash} \cite{gemini} (Gemini 3.5) is a closed frontier model that accepts audio natively.

\vspace{-0.5em}\subsection{Cascaded systems}
\begin{table*}[t!]
\centering
\caption{Comparing audio captioning models across three datasets.
Metrics averaged across eight text models in \Cref{ss:llms}.
Shaded columns highlight the headline metrics for the segmentation and description tasks.
The human row is a proxy rather than a ceiling; see \Cref{ss:human}.
}
\label{tbl:audiocap}
\resizebox{\textwidth}{!}{

\begin{tabular}{lc cc >{\columncolor{gray!20}}c cc >{\columncolor{gray!20}}c cc cc >{\columncolor{gray!20}}c cc >{\columncolor{gray!20}}c cc >{\columncolor{gray!20}}c}
\toprule
 & & \multicolumn{8}{c}{\textbf{Ego4D}} & \multicolumn{3}{c}{\textbf{SINS}} & \multicolumn{6}{c}{\textbf{EgoLife}} \\
\cmidrule(lr){3-10} \cmidrule(lr){11-13} \cmidrule(lr){14-19}
 & & \multicolumn{3}{c}{Segmentation} & \multicolumn{3}{c}{Description} & \multicolumn{2}{c}{Utility} & \multicolumn{3}{c}{Segmentation} & \multicolumn{3}{c}{Segmentation} & \multicolumn{3}{c}{Description} \\
\cmidrule(lr){3-5} \cmidrule(lr){6-8} \cmidrule(lr){9-10} \cmidrule(lr){11-13} \cmidrule(lr){14-16} \cmidrule(lr){17-19}
\textbf{Model} & PFLOPs & bF1$\uparrow$ & evER$\downarrow$ & fAcc$\uparrow$ & dR$\uparrow$ & dP$\uparrow$ & dF1$\uparrow$ & Hit$\uparrow$ & Rddc$\downarrow$ & bF1$\uparrow$ & evER$\downarrow$ & fAcc$\uparrow$ & bF1$\uparrow$ & evER$\downarrow$ & fAcc$\uparrow$ & dR$\uparrow$ & dP$\uparrow$ & dF1$\uparrow$ \\
\midrule
EnCLAP          & 6.41  & 0.480 & \textbf{2.528} & 0.320 & 0.222 & 0.164 & 0.188 & 0.151 & 0.349 & 0.444 & \textbf{2.072} & 0.287 & 0.409 & 2.623 & \textbf{0.308} & 0.080 & \textbf{0.504} & 0.138 \\
MSCLAP          & 0.48  & 0.487 & 2.618 & 0.287 & 0.219 & 0.156 & 0.181 & 0.142 & 0.308 & 0.394 & 2.661 & 0.395 & 0.432 & \textbf{2.606} & 0.302 & 0.078 & 0.468 & 0.132 \\
Qwen 2.5-o      & 43.01 & 0.491 & 3.519 & 0.295 & 0.240 & 0.146 & 0.181 & 0.150 & \textbf{0.282} & 0.432 & 2.613 & 0.395 & 0.450 & 4.205 & 0.288 & 0.093 & 0.427 & 0.150 \\
Qwen 3-o        & 6.69  & \textbf{0.502} & 4.444 & 0.313 & \textbf{0.283} & 0.140 & 0.186 & 0.161 & \textbf{0.282} & 0.358 & 3.864 & 0.354 & \textbf{0.460} & 4.877 & 0.295 & \textbf{0.114} & 0.426 & \textbf{0.178} \\
Qwen 3-o Cap.   & 9.24  & 0.450 & 4.499 & 0.283 & 0.271 & 0.136 & 0.173 & 0.144 & 0.294 & 0.289 & 4.312 & 0.255 & 0.442 & 5.052 & 0.277 & 0.111 & 0.429 & 0.172 \\
AF 3            & 54.48 & 0.488 & 2.757 & \textbf{0.330} & 0.245 & \textbf{0.166} & \textbf{0.197} & \textbf{0.165} & 0.320 & \textbf{0.468} & 2.303 & \textbf{0.437} & 0.437 & 3.134 & 0.293 & 0.088 & 0.503 & 0.149 \\
\midrule
Human           & ---   & 0.460 & 0.745 & 0.646 & 0.550 & 0.550 & 0.550 & 0.329 & 0.227 &   ---   &   ---   &   ---   &   ---   &   ---   &   ---   &   ---   &   ---   &   ---   \\
\bottomrule
\end{tabular}
}
\vspace{-1em}
\end{table*}
\begin{table*}[t!]
\centering
\caption{Comparing text models across three datasets. Metrics averaged across six audio captioning models in \Cref{ss:captioners}.}
\label{tbl:llm}
\resizebox{\textwidth}{!}{
\begin{tabular}{lc cc >{\columncolor{gray!20}}c cc >{\columncolor{gray!20}}c cc cc >{\columncolor{gray!20}}c cc >{\columncolor{gray!20}}c cc >{\columncolor{gray!20}}c}
\toprule
 & & \multicolumn{8}{c}{\textbf{Ego4D}} & \multicolumn{3}{c}{\textbf{SINS}} & \multicolumn{6}{c}{\textbf{EgoLife}} \\
\cmidrule(lr){3-10} \cmidrule(lr){11-13} \cmidrule(lr){14-19}
 & & \multicolumn{3}{c}{Segmentation} & \multicolumn{3}{c}{Description} & \multicolumn{2}{c}{Utility} & \multicolumn{3}{c}{Segmentation} & \multicolumn{3}{c}{Segmentation} & \multicolumn{3}{c}{Description} \\
\cmidrule(lr){3-5} \cmidrule(lr){6-8} \cmidrule(lr){9-10} \cmidrule(lr){11-13} \cmidrule(lr){14-16} \cmidrule(lr){17-19}
\textbf{Model} & PFLOPs & bF1$\uparrow$ & evER$\downarrow$ & fAcc$\uparrow$ & dR$\uparrow$ & dP$\uparrow$ & dF1$\uparrow$ & Hit$\uparrow$ & Rddc$\downarrow$ & bF1$\uparrow$ & evER$\downarrow$ & fAcc$\uparrow$ & bF1$\uparrow$ & evER$\downarrow$ & fAcc$\uparrow$ & dR$\uparrow$ & dP$\uparrow$ & dF1$\uparrow$ \\
\midrule
K2-V2      & 62.11 & 0.404 & \textbf{1.773} & 0.117 & 0.107 & 0.106 & 0.101 & 0.061 & \textbf{0.223} & 0.244 & \textbf{1.277} & 0.186 & 0.322 & \textbf{1.526} & 0.117 & 0.035 & 0.402 & 0.063 \\
OLMo 3.1   & 18.77 & 0.496 & 3.685 & 0.179 & 0.221 & 0.157 & 0.184 & 0.139 & 0.331 & 0.332 & 3.339 & 0.219 & \textbf{0.498} & 4.685 & 0.170 & 0.100 & 0.483 & 0.165 \\
Qwen 2.5   & 39.60 & 0.492 & 3.361 & 0.367 & 0.270 & 0.141 & 0.183 & 0.172 & 0.282 & 0.423 & 3.484 & 0.337 & 0.436 & 3.683 & 0.352 & 0.099 & 0.448 & 0.160 \\
Llama 3.3  & 41.36 & \textbf{0.500} & 4.565 & 0.363 & 0.272 & 0.151 & 0.193 & 0.162 & 0.351 & 0.332 & 4.688 & 0.331 & 0.471 & 4.724 & \textbf{0.369} & 0.111 & 0.457 & 0.178 \\
Qwen 3     & 19.70 & 0.481 & 3.472 & 0.308 & 0.271 & 0.148 & 0.190 & 0.162 & 0.328 & 0.387 & 3.499 & 0.337 & 0.451 & 4.086 & 0.345 & 0.107 & 0.444 & 0.172 \\
Gemma 4    & 16.93 & \textbf{0.500} & 3.511 & 0.370 & 0.271 & \textbf{0.189} & \textbf{0.219} & 0.166 & 0.293 & 0.435 & 3.103 & 0.442 & 0.444 & 3.682 & 0.326 & 0.095 & \textbf{0.504} & 0.157 \\
Gemini 3.5 & ---   & 0.492 & 3.096 & \textbf{0.373} & 0.271 & 0.156 & 0.197 & 0.175 & 0.317 & 0.507 & 1.927 & \textbf{0.513} & 0.441 & 3.337 & 0.334 & 0.093 & 0.467 & 0.153 \\
GPT 5.6    & ---   & 0.499 & 3.691 & 0.361 & \textbf{0.289} & 0.163 & 0.205 & \textbf{0.180} & 0.319 & \textbf{0.518} & 2.450 & 0.467 & 0.446 & 4.276 & 0.336 & \textbf{0.112} & 0.473 & \textbf{0.179} \\
\midrule
Human      & ---   & 0.460 & 0.745 & 0.646 & 0.550 & 0.550 & 0.550 & 0.329 & 0.227 &   ---   &   ---   &   ---   &   ---   &   ---   &   ---   &   ---   &   ---   &   ---   \\
\bottomrule
\end{tabular}
}
\vspace{-1em}
\end{table*}
\noindent\textbf{Audio captioning models.}\label{ss:captioners}
The three open E2E models (Qwen 2.5-o, Qwen 3-o, and AF 3) double as captioners.
We add Qwen 3-Omni-30B-A3B-Captioner (Qwen 3-o Cap.), the caption-tuned variant of Qwen 3-o, and two CLAP models, \textit{EnCLAP-large} \cite{kim2024enclap} and \textit{MSCLAP-2023} \cite{elizalde2024natural} (\texttt{clapcap} version), motivated by the finding that CLAP can outperform larger audio-language models at audio pairwise comparison task \cite{guo2026brace}.

\noindent\textbf{Text models.}\label{ss:llms}
We test eight instruction-tuned text models.
Fully open (data, code, and weights): \textit{OLMo 3.1-32B-Instruct} \cite{olmo2025olmo} (OLMo 3.1) and \textit{K2-V2-70B-Instruct} \cite{liu2025k2} (K2-V2).
Open weights, following the SPAN temporal reasoning benchmark \cite{miao2026span}: \textit{Qwen 2.5-72B-Instruct} \cite{yang2025qwen2} (Qwen 2.5) and \textit{Llama 3.3-70B-Instruct} \cite{grattafiori2024llama} (Llama 3.3).
Frontier open weights, sized comparably to OLMo 3.1: \textit{Qwen 3-32B} \cite{yang2025qwen3} (Qwen 3) and \textit{Gemma 4-31B-it} \cite{team2026gemma} (Gemma 4).
Closed APIs, lightweight tier: \textit{Gemini 3.5 Flash} \cite{gemini} (Gemini 3.5) and \textit{GPT 5.6 Luna} \cite{gpt} (GPT 5.6).

\vspace{-0.5em}
\section{Results}\label{s:results}\vspace{-0.5em}
\noindent\textbf{Settings.}
E2E models receive ten-minute windows, the longest AF 3 accepts.
Captioners receive ten-second windows, matching the AudioSet format \cite{gemmeke2017audio}.
The captions are then grouped into ten minutes for the text model, for a fair comparison to E2E.
Each window is processed independently, decoded greedily, and parsed leniently \cite{jsonrepair}, except for CLAP models, which use beam search as default.
Boundaries are predicted at a minute-level resolution.

\subsection{Comparing different components of cascaded systems}\label{ss:result-casc}
A cascaded system pairs one captioner with one text model, total of $6 \times 8 = 48$ systems.
\Cref{tbl:audiocap} averages over the text models for each captioner, and vice versa for \Cref{tbl:llm}.
Full results are in \Cref{apdx:full}.

Among captioners in \Cref{tbl:audiocap}, AF 3 leads on Ego4D and SINS, though it is also the most computationally heavy.
MSCLAP is not the best, but comes close to AF 3 at a fraction of the cost.
Within the Qwen family, Qwen 3-o is best everywhere except SINS, and interestingly, beats its own captioner-tuned variant throughout.

Among text models (\Cref{tbl:llm}), closed APIs are often the best, where Gemma 4 is generally the strongest non-API model.
Additionally, K2-V2's low evER is an artifact of returning no parseable output for 64\% of the windows.
As such, for all the below experiments, we use AF 3 as the captioner, and compare three text models, Gemini 3.5, GPT 5.6, and Gemma 4, for the cascaded systems.

\subsection{Comparing cascaded systems with E2E systems}\label{ss:casc-vs-e2e}
\begin{table*}[t!]
\centering
\caption{Comparing cascaded and E2E systems across three datasets. Cascaded systems use AF 3 as the audio captioning model.
Shaded columns highlight the headline metrics for the segmentation and description tasks.
The human row is a proxy rather than a ceiling; see \Cref{ss:human}.}
\label{tbl:e2e}
\resizebox{\textwidth}{!}{
\begin{tabular}{lc cc >{\columncolor{gray!20}}c cc >{\columncolor{gray!20}}c cc cc >{\columncolor{gray!20}}c cc >{\columncolor{gray!20}}c cc >{\columncolor{gray!20}}c}
\toprule
 & & \multicolumn{8}{c}{\textbf{Ego4D}} & \multicolumn{3}{c}{\textbf{SINS}} & \multicolumn{6}{c}{\textbf{EgoLife}} \\
\cmidrule(lr){3-10} \cmidrule(lr){11-13} \cmidrule(lr){14-19}
 & & \multicolumn{3}{c}{Segmentation} & \multicolumn{3}{c}{Description} & \multicolumn{2}{c}{Utility} & \multicolumn{3}{c}{Segmentation} & \multicolumn{3}{c}{Segmentation} & \multicolumn{3}{c}{Description} \\
\cmidrule(lr){3-5} \cmidrule(lr){6-8} \cmidrule(lr){9-10} \cmidrule(lr){11-13} \cmidrule(lr){14-16} \cmidrule(lr){17-19}
\textbf{System} & PFLOPs & bF1$\uparrow$ & evER$\downarrow$ & fAcc$\uparrow$ & dR$\uparrow$ & dP$\uparrow$ & dF1$\uparrow$ & Hit$\uparrow$ & Rddc$\downarrow$ & bF1$\uparrow$ & evER$\downarrow$ & fAcc$\uparrow$ & bF1$\uparrow$ & evER$\downarrow$ & fAcc$\uparrow$ & dR$\uparrow$ & dP$\uparrow$ & dF1$\uparrow$ \\
\midrule
Casc. Gemini 3.5 & --- & 0.493 & 2.319 & 0.397 & 0.276 & 0.175 & 0.215 & 0.198 & 0.340 & 0.612 & 1.133 & \textbf{0.571} & 0.429 & 2.686 & 0.325 & 0.087 & 0.512 & 0.149 \\
Casc. GPT 5.6 & --- & 0.500 & 2.863 & 0.383 & 0.273 & 0.182 & 0.218 & 0.191 & 0.340 & \textbf{0.640} & 1.284 & 0.527 & 0.428 & 3.429 & 0.327 & \textbf{0.096} & 0.510 & \textbf{0.162} \\
Casc. Gemma 4 & 71.41 & \textbf{0.508} & 2.608 & 0.398 & 0.262 & 0.218 & 0.238 & 0.160 & 0.304 & 0.547 & 1.588 & 0.512 & \textbf{0.430} & 2.822 & 0.311 & 0.085 & 0.560 & 0.147 \\
\quad $+$ SFT & 71.12 & 0.247 & 0.785 & 0.520 & 0.316 & 0.403 & 0.355 & 0.180 & 0.207 & 0.573 & \textbf{0.668} & 0.421 & 0.254 & \textbf{0.756} & 0.335 & 0.044 & 0.706 & 0.084 \\
\midrule
E2E Gemini 3.5 & --- & 0.437 & 1.357 & 0.445 & 0.271 & 0.209 & 0.236 & 0.214 & 0.289 & 0.368 & 2.256 & 0.495 & 0.422 & 1.648 & 0.331 & 0.087 & 0.501 & 0.149 \\
E2E Qwen 3-o & 11.97 & 0.349 & 1.344 & 0.413 & 0.214 & 0.161 & 0.183 & 0.191 & 0.347 & 0.415 & 1.950 & 0.426 & 0.334 & 1.469 & 0.273 & 0.082 & 0.463 & 0.139 \\
\quad + SFT & 11.67 & 0.252 & \textbf{0.681} & \textbf{0.633} & \textbf{0.376} & \textbf{0.481} & \textbf{0.422} & \textbf{0.254} & 0.212 & 0.285 & 0.721 & 0.483 & 0.241 & 0.770 & 0.332 & 0.057 & \textbf{0.773} & 0.107 \\
E2E Qwen 2.5-o & 22.71 & 0.359 & 1.404 & 0.398 & 0.100 & 0.306 & 0.151 & 0.131 & 0.389 & 0.289 & 2.980 & 0.275 & 0.349 & 0.913 & \textbf{0.638} & 0.056 & 0.643 & 0.103 \\
E2E AF 3 & 51.51 & 0.000 & 1.000 & 0.000 & 0.000 & --- & --- & 0.000 & \textbf{0.000} & 0.000 & 1.000 & 0.000 & 0.000 & 1.000 & 0.000 & 0.000 & --- & --- \\
\midrule
Human & --- & 0.460 & 0.745 & 0.646 & 0.550 & 0.550 & 0.550 & 0.329 & 0.227 & --- & --- & --- & --- & --- & --- & --- & --- & --- \\
\bottomrule
\end{tabular}
}
\vspace{-1em}
\end{table*}

We now compare the three best cascaded systems from \Cref{ss:result-casc} with four E2E systems in \Cref{tbl:e2e} on headline metrics, fAcc and dF1.
E2E Gemini 3.5 is the best on Ego4D and EgoLife across both systems, beating cascaded Gemini 3.5 on both.
SINS is the exception, as the three cascade systems lead the four E2E systems.
We suspect this might be related to SINS being longer (\Cref{tbl:dataset}); \Cref{ss:window} examines how systems respond to longer input windows.

Among non-API E2E models, Qwen 3-o is the strongest, while AF 3, the best captioner, scores zero as it fails to emit the required output schema.
Also, Qwen 2.5-o's high EgoLife fAcc is due to the model predicting \textit{productive} almost everywhere, and 64.3\% of EgoLife frames are \textit{productive}.
We caution that fAcc rewards the majority class, so we additionally report class-wise accuracy in \Cref{apdx:facc}.

Finally, we emphasize that cascaded systems are not the cheaper option.
Even counting only the text model, Gemma 4 costs 16.93 PFLOPs against 11.97 for E2E Qwen 3-o, and adding the AF 3 captioner brings the cascade to 71.41.

\section{Analysis}
Both cascaded and E2E systems in \Cref{s:results} fall short of the human baseline on Ego4D.
In particular, even the best systems suffer from the \textbf{over-segmentation problem} (\Cref{fig:ft_viz}), showing evER above 100\%, with GPT 5.6 reaching 286\% and 343\% on Ego4D and EgoLife, respectively.
We test three additional methods that may improve the performance: supervised fine-tuning (\Cref{ss:sft}), longer context (\Cref{ss:window}), and reasoning (\Cref{ss:reason})

\subsection{Supervised fine-tuning (SFT)}\label{ss:sft}
\begin{figure*}
    \centering
    \includegraphics[width=0.95\textwidth]{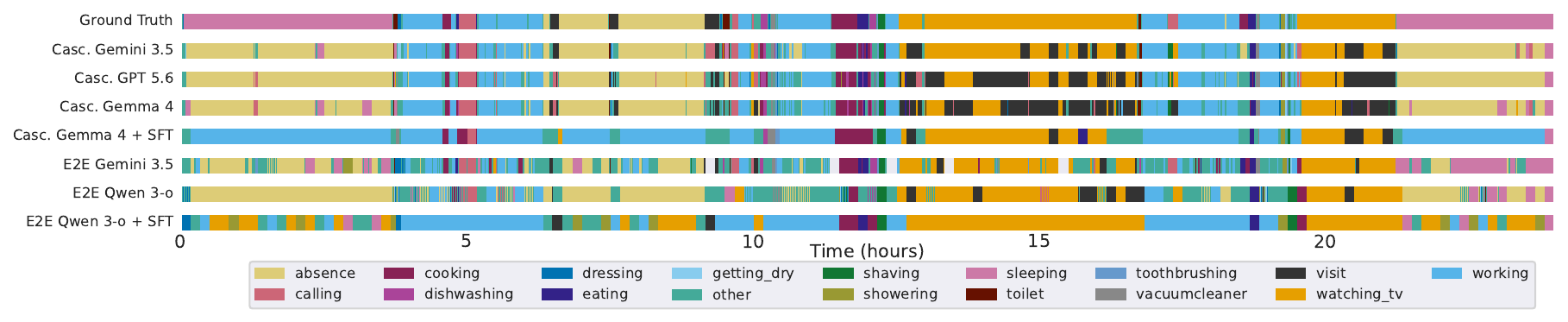}
    \vspace{-1.5em}
    \caption{Visualization of the predicted segmentations for the first 24 hours of SINS dataset using SINS label definition.}
    \label{fig:ft_viz}
    \vspace{-1.5em}
\end{figure*}

As SFT may teach the model both the output format and how finely to segment, we fine-tune Gemma 4 and Qwen 3-o, which are the strongest non-API systems for the cascaded and E2E systems of \Cref{ss:casc-vs-e2e}, on the Ego4D train split, with frozen AF 3 captions as Gemma 4's input.
We employ LoRA \cite{hu2021lora} (r=16, $\alpha$=32, dropout=0.05) for one epoch of training, and use AdamW \cite{loshchilov2017decoupled} at batch size 8 and learning rate 1e-4 with a cosine schedule and 3\% warmup, best-kept by valid loss every 200 steps.

Indeed, \Cref{tbl:e2e} demonstrates that SFT mitigates the over-segmentation problem: evER falls below 100\% for both cascaded Gemma 4 and E2E Qwen 3-o on all datasets, and fAcc sharply improves on Ego4D.
\Cref{fig:ft_viz} further shows that the predicted segmentations become coarser, with the event count falling and mean event length rising on every dataset (average event counts in \Cref{apdx:eventcount}).

However, for SINS and EgoLife, which are out of distribution, the gains are much smaller, and Gemma 4 even degrades on SINS.
One reason is class coverage: SINS consists of 27\% \textit{sleeping} and 17\% \textit{absence}, classes that fine-tuned models avoid because Ego4D contains only events with actions (label distribution in \Cref{apdx:facc}).

Also, EgoLife description distribution differs with Ego4D.
Ego4D averages 3.6 facts per five-minute window against EgoLife's 48.7 (additional statistics in \Cref{apdx:factcount}).
SFT therefore teaches the model to write far less, leading to an increase in dP but a decrease in dR.

Finally, comparing the two best SFT systems, E2E Qwen 3-o improves more: bigger gains on Ego4D and EgoLife, and it alone gains on SINS.
The two gains differ in character. 
Gemma 4 benefits more from learning granularity, reducing its Ego4D event count 4.7$\times$ against 2.1$\times$ for Qwen 3-o (full event count statistics in \Cref{apdx:eventcount}).
Qwen 3-o instead benefits more from better labeling: its \textit{productive} fAcc rises from 27\% to 76\%, against 32\% to 60\% for Gemma 4.

\subsection{Context window size}\label{ss:window}
\begin{figure}
    \centering
    \includegraphics[width=\linewidth]{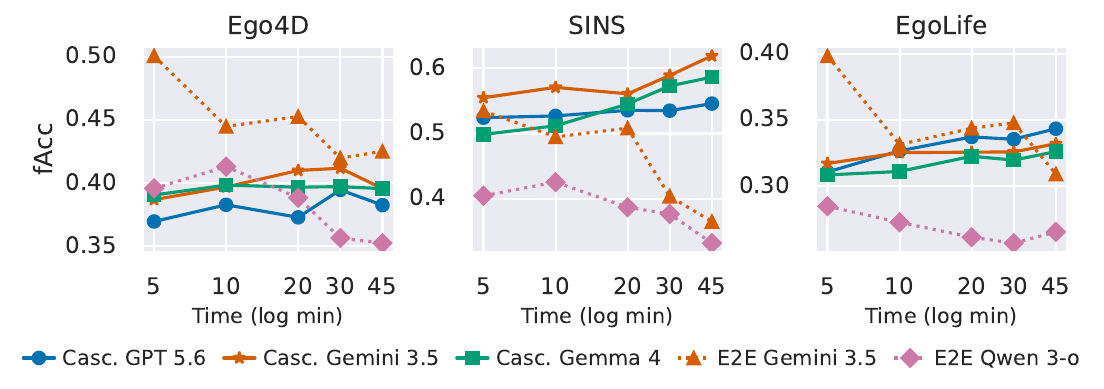}
    \vspace{-2.5em}
    \caption{Segmentation performance ablation on context window size.}
    \label{fig:window}
    \vspace{-1em}
\end{figure}

To test whether longer context helps, we vary the window from 5 to 45 minutes for the top-performing systems of \Cref{ss:casc-vs-e2e}.
\Cref{fig:window} shows that segmentation quality for cascaded systems is similar or better at longer windows, whereas for E2E systems it peaks at 5 or 10 minutes and then declines.
The results suggest that cascaded systems can exploit longer context better than E2E systems.
Additionally, description precision tends to rise while recall falls, as predicted length of each segment becomes longer (average length statistics in \Cref{apdx:desc}).

\subsection{Reasoning}\label{ss:reason}
We enable reasoning capabilities of the topline models.
Following \cite{muennighoff2025s1}, we use hard budget caps of 1024, 2048, and 4096 tokens for Qwen 3-o and Gemma 4, where we use the Thinker variant for Qwen 3-o.
We use the same values as Gemini's soft cap, and low, medium, and high reasoning effort for GPT 5.6.

\begin{figure}
    \centering
    \includegraphics[width=\linewidth]{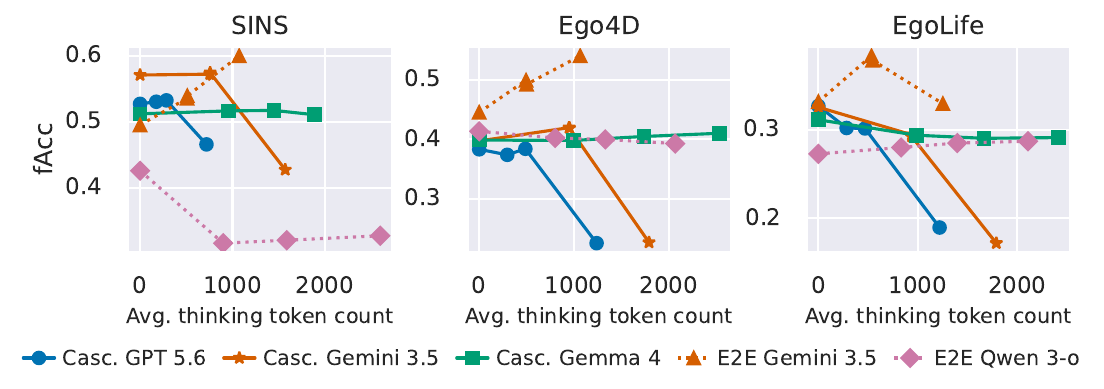}
    \vspace{-2.5em}
    \caption{Segmentation performance ablation on reasoning budget.}
    \label{fig:think}
    \vspace{-1em}
\end{figure}

As shown in \Cref{fig:think}, only E2E Gemini benefits, its segmentation performance rises monotonically with budget.
Every other system is similar or worse.
Much of this is explained by the models' failure to emit the required output format.
E2E Gemini's gain coincides with correcting its parsing failures from 15.4\% to nearly none (additional parsing failure statistics in \Cref{apdx:thinkfailure}).
The drops for cascaded Gemini 3.5 and GPT 5.6 at the highest budget come from output truncation, as reasoning exhausts the output budget.
Cascaded Gemma 4 and E2E Qwen 3-o already parse well, and more thinking slightly worsens or does not change it, leading to no gains.
The trend is the same for all the other datasets.
The results suggest reasoning helps instruction following rather than audio understanding.

\vspace{-0.5em}\section{Conclusion}\vspace{-0.5em}
We introduce the \textsc{Logbook} benchmark, designing the task of long-form audio understanding as a gap-free segmentation, where each segment carries an event label and a description.
We demonstrate that the task, even without vision, is feasible with currently available models.
However, all systems over-segment, which SFT mitigates.
Also, we show that E2E can outperform cascaded systems on both headline metrics and efficiency, yet degrade with longer windows.
Finally, we show that reasoning yields limited gains by improving instruction following rather than audio understanding.
Future works can be done on additional event vocabularies, structured decoding, reinforcement learning, long context for E2E models, and assistive uses for deaf and hard-of-hearing users or personalized assistants.

\section{Compliance with Ethical Standards}\vspace{-0.5em}
This study relied entirely on publicly available datasets and models.
Consequently, ethical approval was not required.

\section{Acknowledgments}\vspace{-0.5em}
We thank Ray Huang and Gautam Tiwari for valuable discussions.

\bibliographystyle{IEEEbib}
\bibliography{strings,refs}

@article{grauman2024ego4d,
  title={Ego4d: Around the world in 3,000 hours of egocentric video},
  author={Grauman, Kristen and others},
  journal={IEEE TPAMI},
  year={2024},
}

@inproceedings{lin2026wearvox,
  title={Wearvox: An egocentric multichannel voice assistant benchmark for wearables},
  author={Lin, Zhaojiang and others},
  booktitle={ICLR},
  year={2026}
}

@inproceedings{yang2025egolife,
  title={Egolife: Towards egocentric life assistant},
  author={Yang, Jingkang and others},
  booktitle={CVPR},
  year={2025},
}

@inproceedings{hu2021lora,
  title={Lora: Low-rank adaptation of large language models},
  author={Hu, Edward J and others},
  booktitle={ICLR},
  year={2022}
}

@inproceedings{loshchilov2017decoupled,
  title={Decoupled weight decay regularization},
  author={Loshchilov, Ilya and Hutter, Frank},
  booktitle={ICLR},
  year={2019}
}

@article{paruchuri2025egotrigger,
  title={Egotrigger: Toward audio-driven image capture for human memory enhancement in all-day energy-efficient smart glasses},
  author={Paruchuri, Akshay and others},
  journal={IEEE TVCG},
  year={2025},
}

@article{ahia2025blab,
  title={Blab: Brutally long audio bench},
  author={Ahia, Orevaoghene and others},
  journal={arXiv preprint 2505.03054},
  year={2025}
}

@article{luo2026chronosaudio,
  title={Chronosaudio: A comprehensive long-audio benchmark for evaluating audio-large language models},
  author={Luo, Kaiwen and others},
  journal={arXiv preprint 2601.04876},
  year={2026}
}

@article{ghosh2026audio,
  title={Audio flamingo 3: Advancing audio intelligence with fully open large audio language models},
  author={Ghosh, Sreyan and others},
  journal={NeurIPS},
  year={2026}
}

@article{xu2025qwen3,
  title={Qwen3-omni technical report},
  author={Xu, Jin and others},
  journal={arXiv preprint 2509.17765},
  year={2025}
}

@article{yang2025qwen3,
  title={Qwen3 technical report},
  author={Yang, An and others},
  journal={arXiv preprint 2505.09388},
  year={2025}
}

@article{dekkers2017sins,
  title={The SINS database for detection of daily activities in a home environment using an acoustic sensor network},
  author={Dekkers, Gert and others},
  journal={DCASE},
  year={2017},
}

@inproceedings{gemmeke2017audio,
  title={Audio set: An ontology and human-labeled dataset for audio events},
  author={Gemmeke, Jort F and others},
  booktitle={ICASSP},
  year={2017},
}

@article{kaplan2020scaling,
  title={Scaling laws for neural language models},
  author={Kaplan, Jared and others},
  journal={arXiv preprint 2001.08361},
  year={2020}
}

@inproceedings{min2023factscore,
    title = "{FA}ct{S}core: Fine-grained Atomic Evaluation of Factual Precision in Long Form Text Generation",
    author = "Min, Sewon  and others",
    booktitle = "EMNLP",
    year = "2023",
}

@article{lage2025openfactscore,
  title={Openfactscore: Open-source atomic evaluation of factuality in text generation},
  author={Lage, Lucas Fonseca and Ostermann, Simon},
  journal={arXiv preprint 2507.05965},
  year={2025}
}

@article{fredianelli2025environmental,
  title={Environmental noise dataset for sound event classification and detection},
  author={Fredianelli, Luca and others},
  journal={Scientific Data},
  year={2025},
}

@inproceedings{walsh20252,
  title={2 OLMo 2 furious},
  author={Walsh, Evan Pete and others},
  booktitle={COLM},
  year={2025}
}

@inproceedings{lewis2020bart,
  title={BART: Denoising sequence-to-sequence pre-training for natural language generation, translation, and comprehension},
  author={Lewis, Mike and others},
  booktitle={ACL},
  year={2020}
}

@article{costa2022no,
  title={No language left behind: Scaling human-centered machine translation},
  author={Costa-Juss{\`a}, Marta R and others},
  journal={arXiv preprint 2207.04672},
  year={2022}
}

@article{xu2025qwen25,
  title={Qwen2.5-Omni Technical Report},
  author={Jin Xu and others},
  journal={arXiv preprint 2503.20215},
  year={2025}
}

@article{team2026gemma,
  title={Gemma 4 technical report},
  author={Team, Gemma and others},
  journal={arXiv preprint 2607.02770},
  year={2026}
}

@article{yang2025qwen2,
  title={Qwen2.5 technical report},
  author={An Yang and others},
  journal={arXiv preprint 2412.15115},
  year={2025}
}

@inproceedings{radford2023robust,
  title={Robust speech recognition via large-scale weak supervision},
  author={Radford, Alec and others},
  booktitle={ICML},
  year={2023},
}

@software{jsonrepair,
    author  = "Stefano {Baccianella}",
    month   = "feb",
    title   = "JSON Repair: A python module to repair invalid JSON",
    url     = "https://github.com/mangiucugna/json_repair",
    version = "0.39.1",
    year    = 2025
}

@inproceedings{DrossosLV20,
  author       = {Konstantinos Drossos and others},
  title        = {Clotho: an Audio Captioning Dataset},
  booktitle    = {ICASSP},
  year         = {2020},
}

@article{damen2020epic,
  title={The epic-kitchens dataset: Collection, challenges and baselines},
  author={Damen, Dima and others},
  journal={IEEE TPAMI},
  year={2020},
}

@inproceedings{perrett2025hd,
  title={{HD-EPIC}: A highly-detailed egocentric video dataset},
  author={Perrett, Toby and others},
  booktitle={CVPR},
  year={2025},
}

@inproceedings{kitani2011fast,
  title={Fast unsupervised ego-action learning for first-person sports videos},
  author={Kitani, Kris M and others},
  booktitle={CVPR},
  year={2011},
}

@inproceedings{aghazadeh2011novelty,
  title={Novelty detection from an ego-centric perspective},
  author={Aghazadeh, Omid and others},
  booktitle={CVPR},
  year={2011},
}

@article{youhome2022,
  title={YouHome System and Dataset: Making Your Home Know You Better},
  author={Pan, Junhao and others},
  journal={IEEE iSES},
  year={2022}
}

@article{mesaros2021sound,
  title={Sound event detection: A tutorial},
  author={Mesaros, Annamaria and others},
  journal={IEEE SPM},
  year={2021},
}

@article{mei2022automated,
  title={Automated audio captioning: An overview of recent progress and new challenges},
  author={Mei, Xinhao and others},
  journal={Journal on Audio, Speech, and Music Processing},
  year={2022},
}

@article{guo2026brace,
  title={Brace: A benchmark for robust audio caption quality evaluation},
  author={Guo, Tianyu and others},
  journal={NeurIPS},
  year={2026}
}

@inproceedings{elizalde2024natural,
  title={Natural language supervision for general-purpose audio representations},
  author={Elizalde, Benjamin and others},
  booktitle={ICASSP},
  year={2024},
}

@inproceedings{kim2024enclap,
  title={Enclap: Combining neural audio codec and audio-text joint embedding for automated audio captioning},
  author={Kim, Jaeyeon and others},
  booktitle={ICASSP},
  year={2024},
}

@article{liu2025k2,
  title={K2-v2: A 360-open, reasoning-enhanced llm},
  author={Liu, Zhengzhong and others},
  journal={arXiv preprint 2512.06201},
  year={2025}
}

@article{grattafiori2024llama,
  title={The llama 3 herd of models},
  author={Grattafiori, Aaron and others},
  journal={arXiv preprint 2407.21783},
  year={2024}
}

@inproceedings{miao2026span,
  title={SPAN: Benchmarking and Improving Cross-Calendar Temporal Reasoning of Large Language Models},
  author={Miao, Zhongjian and others},
  booktitle={AAAI},
  year={2026}
}

@misc{gemini,
  author       = {{Google}},
  title        = {Gemini 3.5 Flash},
  year         = {2026},
  howpublished = {\url{https://ai.google.dev/gemini-api/docs/models/gemini-3.5-flash}}
}

@misc{gpt,
  author       = {{OpenAI}},
  title        = {GPT-5.6 Luna},
  year         = {2026},
  howpublished = {\url{https://developers.openai.com/api/docs/models/gpt-5.6-luna}},
}

@inproceedings{muennighoff2025s1,
  title={s1: Simple test-time scaling},
  author={Muennighoff, Niklas and others},
  booktitle={EMNLP},
  year={2025}
}

@article{shao2026listening,
  title={Listening with Time: Precise Temporal Awareness for Long-Form Audio Understanding},
  author={Shao, Mingchen and others},
  journal={arXiv preprint 2604.22245},
  year={2026}
}

@techreport{bls2026atus,
  author      = {{U.S. Bureau of Labor Statistics}},
  title       = {American Time Use Survey},
  institution = {U.S. Bureau of Labor Statistics},
  year        = {2026},
  month       = {June},
  day         = {25},
  url         = {https://www.bls.gov/news.release/pdf/atus.pdf}
}

@article{olmo2025olmo,
  title={Olmo 3},
  author={Olmo, Team and others},
  journal={arXiv preprint 2512.13961},
  year={2025}
}

@inproceedings{ShazeerMMDLHD17,
  author       = {Noam Shazeer and others},
  title        = {Outrageously Large Neural Networks: The Sparsely-Gated Mixture-of-Experts
                  Layer},
  booktitle    = {ICLR},
  year         = {2017},
}

@inproceedings{serizel2020sound,
  title={Sound event detection in synthetic domestic environments},
  author={Serizel, Romain and others},
  booktitle={ICASSP},
  year={2020},
}

\newpage
\appendix
\crefalias{section}{appendix}
\crefalias{subsection}{appendix}
\section{Appendix}\label{apdx}

\subsection{Ablation on context window size for description quality}\label{apdx:desc}
\begin{figure}[t]
    \centering
    \includegraphics[width=0.7\linewidth]{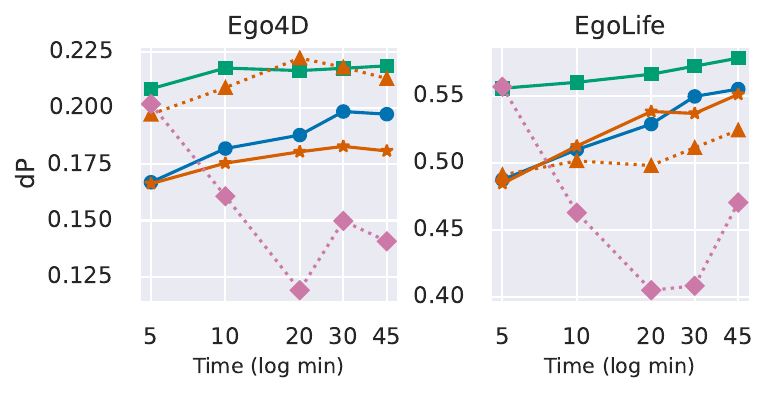}
    \includegraphics[width=0.7\linewidth]{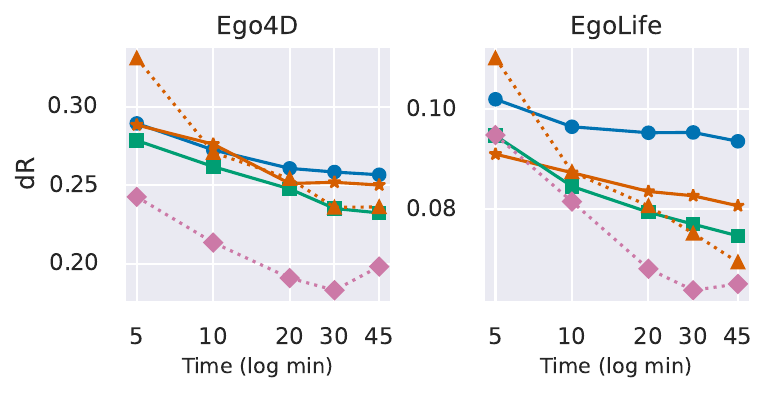}
    \includegraphics[width=\linewidth]{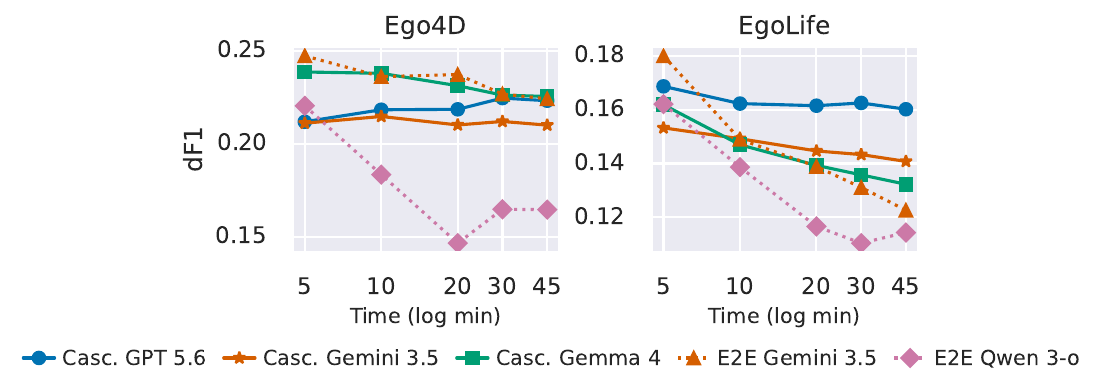}
    \caption{Description performance ablation on the context window size.}
    \label{fig:window_desc}
\end{figure}

\begin{table}[t]
\centering
\caption{Events per 5-minute interval across Ego4D, EgoLife, and SINS datasets.
Average counts per five minutes, and average length for each event in seconds.
}
\resizebox{\linewidth}{!}{
\begin{tabular}{lccccccc}
\toprule
 & & \multicolumn{2}{c}{\textbf{Ego4D}} & \multicolumn{2}{c}{\textbf{EgoLife}} & \multicolumn{2}{c}{\textbf{SINS}} \\
\cmidrule(lr){3-4} \cmidrule(lr){5-6} \cmidrule(lr){7-8}
\textbf{System} & \textbf{Window} & \textbf{Cnt.} & \textbf{Len.} & \textbf{Cnt.} & \textbf{Len.} & \textbf{Cnt.} & \textbf{Len.} \\
\midrule
\multirow{5}{*}{Casc. Gemini 3.5} & 5min & 1.84 & 162 & 1.57 & 191 & 0.88 & 341 \\
 & 10min & 1.26 & 236 & 1.12 & 267 & 0.57 & 525 \\
 & 20min & 1.08 & 275 & 0.90 & 330 & 0.43 & 695 \\
 & 30min & 1.05 & 282 & 0.90 & 332 & 0.43 & 697 \\
 & 45min & 1.05 & 283 & 0.90 & 328 & 0.46 & 659 \\
\midrule
\multirow{5}{*}{Casc. GPT 5.6} & 5min & 1.89 & 158 & 1.58 & 190 & 0.87 & 343 \\
 & 10min & 1.51 & 195 & 1.31 & 223 & 0.63 & 474 \\
 & 20min & 1.40 & 211 & 1.12 & 268 & 0.51 & 588 \\
 & 30min & 1.31 & 228 & 1.10 & 272 & 0.43 & 705 \\
 & 45min & 1.26 & 232 & 0.97 & 309 & 0.44 & 677 \\
\midrule
\multirow{5}{*}{Casc. Gemma 4} & 5min & 1.83 & 164 & 1.53 & 196 & 1.04 & 289 \\
 & 10min & 1.40 & 214 & 1.17 & 256 & 0.73 & 412 \\
 & 20min & 1.25 & 240 & 1.03 & 291 & 0.66 & 455 \\
 & 30min & 1.08 & 274 & 0.95 & 315 & 0.63 & 478 \\
 & 45min & 1.08 & 276 & 0.83 & 359 & 0.59 & 509 \\
\midrule
\multirow{5}{*}{E2E Gemini 3.5} & 5min & 0.94 & 305 & 1.02 & 281 & 1.38 & 216 \\
 & 10min & 0.66 & 384 & 0.64 & 384 & 0.95 & 307 \\
 & 20min & 0.54 & 466 & 0.54 & 464 & 0.79 & 368 \\
 & 30min & 0.49 & 506 & 0.47 & 517 & 0.75 & 367 \\
 & 45min & 0.46 & 543 & 0.39 & 602 & 0.66 & 418 \\
\midrule
\multirow{5}{*}{E2E Qwen 3-o} & 5min & 0.92 & 324 & 0.91 & 328 & 1.46 & 205 \\
 & 10min & 0.60 & 500 & 0.49 & 606 & 0.83 & 360 \\
 & 20min & 0.56 & 533 & 0.70 & 425 & 1.04 & 289 \\
 & 30min & 0.47 & 632 & 0.72 & 415 & 0.55 & 538 \\
 & 45min & 0.25 & 1175 & 0.23 & 1296 & 0.16 & 1797 \\
\midrule
GT & -- & 0.57 & 528 & 0.40 & 754 & 0.32 & 952 \\
\bottomrule
\end{tabular}
}%
\label{tbl:eventwindow}
\end{table}
\Cref{tbl:eventwindow} reports the average number of predicted events per five minutes and their average length.
Longer input windows produce longer predicted events across every system, and correspondingly fewer facts.
\Cref{fig:window_desc} shows the effect on description quality: precision generally rises while recall falls, and the fall in recall dominates, so dF1 decreases overall.

\subsection{Ablation on reasoning budget for description quality}\label{apdx:desc-think}
\begin{figure}[t]
    \centering
    \includegraphics[width=0.98\linewidth]{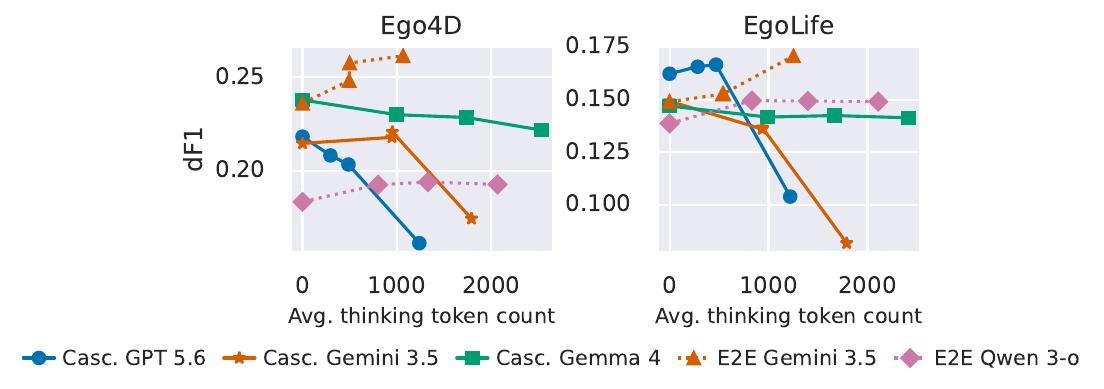}
    \caption{Description performance ablation on the reasoning budget.}
    \label{fig:think_desc}
\end{figure}

\Cref{fig:think_desc} reports description quality with respect to reasoning budget.
It follows nearly the same trend as segmentation in \Cref{fig:think}: E2E Gemini 3.5 improves, cascaded Gemini 3.5 and GPT 5.6 collapse at the largest budget, and the rest are flat.
Since a window that fails to parse yields neither segments nor facts, this consistency further supports attributing the changes to parsing failures rather than to improved understanding.

\subsection{Fact statistics}\label{apdx:factcount}
\Cref{tbl:factcount} reports the average number of atomic facts per five-minute interval.
Where every system produces a similar amount of facts on both datasets, the ground truth differing more than thirteenfold (3.64 on Ego4D against 48.67 on EgoLife).
SFT roughly halves it, bringing it close to the Ego4D reference it was trained on and further from EgoLife's.

\subsection{Event Statistics}\label{apdx:eventcount}
\Cref{tbl:eventcount} reports predicted event counts and lengths before and after SFT. 
GT is calculated only with the first annotator for Ego4D.
SFT cuts the Ego4D event count 4.7$\times$ for the cascade and 2.1$\times$ for the E2E system, in both cases past the reference: both settle near 0.3 events per five minutes, with the same pattern on EgoLife and SINS.

\subsection{Parsing Failure Statistics}\label{apdx:thinkfailure}
\Cref{tbl:thinkfailure} reports the fraction of output predictions whose output could not be parsed into the required schema.
Two model-side failures that often arise: outputs that terminate normally but do not follow the schema, or outputs truncated when the budget is exhausted.
At baseline, E2E Gemini's outputs terminate normally yet do not follow the schema; reasoning corrects this and its failure rate falls from 15.4\% to 0.6\% on Ego4D.
At the largest budget, cascaded Gemini and GPT 5.6 instead exhaust their completion budget on reasoning tokens and their outputs are truncated, pushing failures to 49.2\% and 43.1\%.

\begin{table}[t]
\caption{Facts per 5-minute interval for Ego4D and EgoLife datasets. We use first pass annotations for Ego4D to count the facts.}
\label{tbl:factcount}
\centering
\begin{tabular}{lcc}
\toprule
\textbf{System} & \textbf{Ego4D} & \textbf{EgoLife} \\
\midrule
Casc. Gemini 3.5 & 9.17 & 9.21 \\
Casc. GPT 5.6 & 18.75 & 18.94 \\
Casc. Gemma 4 & 9.89 & 10.06 \\
\quad + SFT & 4.62 & 4.91 \\
\midrule
E2E Gemini 3.5 & 8.50 & 10.10 \\
E2E Qwen 3-o & 9.72 & 10.61 \\
\quad + SFT & 4.62 & 5.11 \\
\midrule
GT & 3.64 & 48.67 \\
\bottomrule
\end{tabular}
\end{table}
\begin{table}[h]
\centering
\caption{Events per 5-minute interval across Ego4D, EgoLife, and SINS datasets.
Average counts per five minutes, and average length for each event in seconds.
}
\begin{tabular}{lcccccc}
\toprule
 & \multicolumn{2}{c}{\textbf{Ego4D}} & \multicolumn{2}{c}{\textbf{EgoLife}} & \multicolumn{2}{c}{\textbf{SINS}} \\
\cmidrule(lr){2-3} \cmidrule(lr){4-5} \cmidrule(lr){6-7}
\textbf{System} & \textbf{Cnt.} & \textbf{Len.} & \textbf{Cnt.} & \textbf{Len.} & \textbf{Cnt.} & \textbf{Len.} \\
\midrule
Casc. Gemini 3.5 & 1.26 & 236 & 1.12 & 267 & 0.57 & 525 \\
Casc. GPT 5.6 & 1.51 & 195 & 1.31 & 223 & 0.63 & 474 \\
Casc. Gemma 4 & 1.40 & 214 & 1.17 & 256 & 0.73 & 412 \\
\quad + SFT & 0.30 & 995 & 0.23 & 1296 & 0.21 & 1403 \\
\midrule
E2E Gemini 3.5 & 0.66 & 384 & 0.64 & 384 & 0.95 & 307 \\
E2E Qwen 3-o & 0.60 & 500 & 0.49 & 606 & 0.83 & 360 \\
\quad + SFT & 0.28 & 1051 & 0.24 & 1263 & 0.22 & 1364 \\
\midrule
GT & 0.57 & 528 & 0.40 & 754 & 0.32 & 952 \\
\bottomrule
\end{tabular}
\label{tbl:eventcount}
\end{table}
\begin{table}[h!]
\centering
\caption{Fraction of predictions not following the required format.}
\begin{tabular}{llccc}
\toprule
\textbf{System} & \textbf{Budget} & \textbf{Ego4D} & \textbf{EgoLife} & \textbf{SINS} \\
\midrule
\multirow{4}{*}{Casc. Gemini 3.5} & None & 0.6\% & 0.2\% & 0.0\% \\
 & 1024 & 1.7\% & 3.0\% & 0.9\% \\
 & 2048 & 2.4\% & 2.8\% & 0.9\% \\
 & 4096 & 49.2\% & 48.2\% & 31.1\% \\
\midrule
\multirow{4}{*}{Casc. GPT 5.6} & None & 0.5\% & 0.1\% & 0.0\% \\
 & Low & 1.3\% & 0.9\% & 0.4\% \\
 & Medium & 1.1\% & 0.4\% & 0.1\% \\
 & High & 43.1\% & 38.8\% & 16.1\% \\
\midrule
\multirow{4}{*}{Casc. Gemma 4} & None & 0.4\% & 0.2\% & 0.0\% \\
 & 1024 & 2.1\% & 1.5\% & 0.7\% \\
 & 2048 & 1.2\% & 0.5\% & 0.4\% \\
 & 4096 & 0.5\% & 0.3\% & 0.3\% \\
\midrule
\multirow{4}{*}{E2E Gemini 3.5} & None & 15.4\% & 17.4\% & 3.4\% \\
 & 1024 & 7.4\% & 7.7\% & 2.0\% \\
 & 2048 & 7.5\% & 7.4\% & 2.2\% \\
 & 4096 & 0.6\% & 0.2\% & 0.2\% \\
\midrule
\multirow{4}{*}{E2E Qwen 3-o} & None & 0.3\% & 0.2\% & 0.2\% \\
 & 1024 & 0.5\% & 0.3\% & 0.1\% \\
 & 2048 & 0.3\% & 0.2\% & 0.2\% \\
 & 4096 & 0.3\% & 0.1\% & 0.1\% \\
\bottomrule
\end{tabular}
\label{tbl:thinkfailure}
\end{table}

\newpage
\subsection{Prompts and Class definitions}\label{apdx:prompts}
\begin{promptbox}[title=Prompt for E2E systems]
You are an expert audio analyst for wearable-device recordings of a single person's activities. You receive long audio recordings, to produce a temporal segmentation of the wearer's activities.

# Task
The recording is 10 minutes long. Segment it into non-overlapping, contiguous spans at *whole-minute* boundaries, covering every minute from 0 to 10 using labels from the list below (no gaps). Each segment must span at least one full minute; do not emit sub-minute segments.

# Output format
Return a JSON object with one key, "segments", whose value is an array of objects each with:
  - "start": integer minute (0..10) *relative to the start of this chunk* ("0" = chunk start)
  - "end":   integer minute (0..10) *relative to the start of this chunk*
  - "event": one of the allowed event classes below
  - "description": 1-2 sentences, present tense, naming the concrete actions and objects in this segment. Add detail beyond the event label; infer plausible specifics when evidence is partial. Refer to the recording subject as "the wearer".
Each end must be strictly greater than its start. Spans must be non-overlapping and chronologically ordered. Do not output anything outside the JSON object.

# Allowed event classes (use these labels exactly):
(Different per dataset)
\end{promptbox}

\begin{promptbox}[title=Prompt for Captioners]
Describe the audio clip in one sentence.
\end{promptbox}

\begin{promptbox}[title=Prompt for Text LLMs]
You are an expert audio analyst for wearable-device recordings of a single person's activities. You receive a series of timestamped descriptions, to produce a temporal segmentation of the wearer's activities.

# Task
(Same with E2E)

# Output format
(Same with E2E)

# Allowed event classes (use these labels exactly):
(Same with E2E)

# Audio descriptions
(Output of audio captioners)
\end{promptbox}

\begin{promptbox}[title=Class definition of Ego4D and EgoLife]
- productive: Directed activity that produces output or meets someone's tangible needs: paid or unpaid work (including skilled trades and manual labor as a job), studying or attending class, homework or research, household maintenance (cleaning, laundry, repairs, lawn/garden, pet care), and hands-on dependent care (feeding, dressing, bathing, teaching children or elderly). Exclude: cooking or eating (use `food`); shopping (use `purchasing`); commuting (use `traveling`); recreational or conversational time with family (use `leisure`).
- food: Preparing, eating, or drinking, alone or with others, at home or elsewhere. Includes cooking, setting the table, and kitchen cleanup. Eating in any context belongs here, not in leisure.
- leisure: Activities done for personal enjoyment or relaxation rather than work or output: TV / streaming / movies, music, video games, reading or writing for pleasure, hobbies, arts and crafts, cultural events; exercise and sports of all kinds; personal phone, text, or video calls; in-person socializing or hosting; and casual family time (chatting, playing, reading together). Exclude: household chores or maintenance (cleaning, laundry, repairs, lawn/garden, pet care) and hands-on dependent care (use `productive`); shopping or in-person transactions (use `purchasing`); eating with others (use `food`).
- purchasing: Shopping for goods or services, in-store, online, or by phone, including visits to professionals such as doctors, banks, or hairdressers.
- traveling: Commuting or travel between activities, by car, foot, transit, or any other means. Movement during another activity (e.g., walking around a workshop while working) belongs to that activity.
- other: Any other activities that do not fit the above categories. Also use this when there is no audible activity (silence, ambient quiet, or audio where no clear human activity is occurring).
\end{promptbox}

\begin{promptbox}[title=Class definition of SINS]
- absence: nobody present in the home or in the relevant room; ambient room sound only (HVAC, distant traffic, refrigerator hum).
- calling: phone call - sustained one-sided speech with pauses for the remote party, typically in the living room.
- cooking: kitchen sounds: chopping, frying, pot/pan clatter, microwave, sizzling.
- dishwashing: running water in the sink with dishes clinking and scrubbing - manual washing-up, not a dishwasher cycle.
- dressing: fabric rustling, zippers, drawer/wardrobe sounds, hangers - typically in the bedroom.
- eating: cutlery on plates, chewing, glass clinks, brief speech at the table.
- getting_dry: towel rubbing and brisk fabric-on-skin sounds after bathing.
- other: a person IS present and active but the activity doesn't fit any of the listed categories. Includes transitions between activities, time between entering a room and starting something, sitting on the sofa, or crossing through the hall.
- shaving: electric-razor buzz or wet-shave scraping at the sink.
- showering: continuous water spray, often with bathroom-fan hum.
- sleeping: a person IS present but inactive: very quiet room, slow breathing, occasional rustling. Usually a long, uninterrupted stretch.
- toilet: toilet flush, brief water and seat sounds.
- toothbrushing: rhythmic brushing sound, often with a sink running briefly before or after.
- vacuumcleaner: continuous vacuum motor hum, often moving across the room. Same label regardless of which room.
- visit: multiple voices, conversation, doorbell, footsteps - guests in the home.
- watching_tv: TV speech, music, sound effects; periodic dialogue and music cues, often sustained.
- working: working at a computer: keyboard typing, mouse clicks, occasional brief speech (phone or video calls).
\end{promptbox}

\newpage
\subsection{Full results of cascaded systems}\label{apdx:full}
\begin{table*}[t!]
\centering
\caption{Comparing all the combinations of cascaded systems.}
\label{tbl:casc}
\resizebox{\textwidth}{!}{
\begin{tabular}{lc cc >{\columncolor{gray!20}}c cc >{\columncolor{gray!20}}c cc cc >{\columncolor{gray!20}}c cc >{\columncolor{gray!20}}c cc >{\columncolor{gray!20}}c}
\toprule
 & & \multicolumn{8}{c}{\textbf{Ego4D}} & \multicolumn{3}{c}{\textbf{SINS}} & \multicolumn{6}{c}{\textbf{EgoLife}} \\
\cmidrule(lr){3-10} \cmidrule(lr){11-13} \cmidrule(lr){14-19}
 & & \multicolumn{3}{c}{Segmentation} & \multicolumn{3}{c}{Description} & \multicolumn{2}{c}{Utility} & \multicolumn{3}{c}{Segmentation} & \multicolumn{3}{c}{Segmentation} & \multicolumn{3}{c}{Description} \\
\cmidrule(lr){3-5} \cmidrule(lr){6-8} \cmidrule(lr){9-10} \cmidrule(lr){11-13} \cmidrule(lr){14-16} \cmidrule(lr){17-19}
\textbf{Audio Cap.} & \textbf{Text model} & bF1$\uparrow$ & evER$\downarrow$ & fAcc$\uparrow$ & R$\uparrow$ & P$\uparrow$ & F1$\uparrow$ & Hit$\uparrow$ & rddc$\downarrow$ & bF1$\uparrow$ & evER$\downarrow$ & fAcc$\uparrow$ & bF1$\uparrow$ & evER$\downarrow$ & fAcc$\uparrow$ & R$\uparrow$ & P$\uparrow$ & F1$\uparrow$ \\
\midrule 
\multirow{8}{*}{EnCLAP} & K2-V2 & 0.445 & 1.796 & 0.189 & 0.156 & 0.120 & 0.136 & 0.087 & 0.277 & 0.301 & 1.364 & 0.243 & 0.372 & 1.829 & 0.219 & 0.059 & 0.482 & 0.105 \\
 & OLMo 3.1 & 0.491 & 2.919 & 0.177 & 0.189 & 0.150 & 0.168 & 0.152 & 0.344 & 0.361 & 2.863 & 0.172 & 0.496 & 3.902 & 0.212 & 0.089 & 0.510 & 0.151 \\
 & Qwen 2.5 & 0.481 & 2.375 & 0.374 & 0.238 & 0.163 & 0.194 & 0.176 & 0.337 & 0.497 & 2.229 & 0.185 & 0.381 & 2.354 & 0.330 & 0.078 & 0.518 & 0.135 \\
 & Llama 3.3 & 0.501 & 3.830 & 0.366 & 0.247 & 0.149 & 0.186 & 0.163 & 0.383 & 0.361 & 3.613 & 0.247 & 0.438 & 3.040 & 0.376 & 0.096 & 0.465 & 0.160 \\
 & Qwen 3 & 0.472 & 2.779 & 0.294 & 0.240 & 0.164 & 0.195 & 0.152 & 0.373 & 0.409 & 2.487 & 0.271 & 0.399 & 2.923 & 0.347 & 0.091 & 0.480 & 0.153 \\
 & Gemma 4 & 0.491 & 2.184 & 0.381 & 0.225 & 0.207 & 0.215 & 0.158 & 0.336 & 0.517 & 1.464 & 0.310 & 0.393 & 2.151 & 0.317 & 0.072 & 0.567 & 0.127 \\
 & Gemini 3.5 & 0.484 & 2.039 & 0.398 & 0.242 & 0.176 & 0.204 & 0.154 & 0.369 & 0.547 & 1.204 & 0.479 & 0.401 & 2.197 & 0.329 & 0.076 & 0.520 & 0.133 \\
 & GPT 5.6 & 0.474 & 2.302 & 0.384 & 0.241 & 0.178 & 0.205 & 0.162 & 0.370 & 0.559 & 1.350 & 0.386 & 0.390 & 2.587 & 0.335 & 0.081 & 0.495 & 0.139 \\
\midrule
\multirow{8}{*}{MSCLAP} & K2-V2 & 0.517 & 1.824 & 0.126 & 0.122 & 0.121 & 0.122 & 0.058 & 0.229 & 0.219 & 1.579 & 0.220 & 0.402 & 1.976 & 0.129 & 0.035 & 0.432 & 0.065 \\
 & OLMo 3.1 & 0.458 & 2.356 & 0.139 & 0.191 & 0.137 & 0.160 & 0.136 & 0.301 & 0.308 & 3.091 & 0.198 & 0.508 & 2.761 & 0.124 & 0.085 & 0.459 & 0.144 \\
 & Qwen 2.5 & 0.477 & 2.581 & 0.355 & 0.232 & 0.155 & 0.186 & 0.154 & 0.289 & 0.419 & 3.217 & 0.400 & 0.389 & 2.419 & 0.342 & 0.082 & 0.471 & 0.139 \\
 & Llama 3.3 & 0.515 & 4.387 & 0.348 & 0.260 & 0.156 & 0.194 & 0.147 & 0.366 & 0.313 & 4.984 & 0.336 & 0.458 & 3.766 & 0.383 & 0.100 & 0.452 & 0.164 \\
 & Qwen 3 & 0.446 & 2.576 & 0.229 & 0.241 & 0.167 & 0.197 & 0.156 & 0.323 & 0.366 & 3.277 & 0.402 & 0.446 & 2.256 & 0.288 & 0.089 & 0.459 & 0.149 \\
 & Gemma 4 & 0.488 & 2.422 & 0.377 & 0.226 & 0.194 & 0.209 & 0.160 & 0.308 & 0.459 & 2.043 & 0.563 & 0.419 & 2.443 & 0.374 & 0.072 & 0.526 & 0.126 \\
 & Gemini 3.5 & 0.494 & 2.240 & 0.386 & 0.246 & 0.153 & 0.189 & 0.163 & 0.325 & 0.514 & 1.483 & 0.521 & 0.422 & 2.406 & 0.415 & 0.078 & 0.478 & 0.134 \\
 & GPT 5.6 & 0.502 & 2.558 & 0.340 & 0.233 & 0.163 & 0.192 & 0.163 & 0.319 & 0.552 & 1.615 & 0.517 & 0.416 & 2.821 & 0.361 & 0.080 & 0.467 & 0.137 \\
\midrule
\multirow{8}{*}{Qwen 2.5-o} & K2-V2 & 0.436 & 1.800 & 0.094 & 0.105 & 0.099 & 0.102 & 0.064 & 0.202 & 0.319 & 1.249 & 0.210 & 0.329 & 1.248 & 0.062 & 0.022 & 0.352 & 0.041 \\
 & OLMo 3.1 & 0.510 & 3.595 & 0.155 & 0.205 & 0.142 & 0.168 & 0.136 & 0.281 & 0.352 & 3.224 & 0.249 & 0.506 & 5.170 & 0.150 & 0.096 & 0.445 & 0.159 \\
 & Qwen 2.5 & 0.512 & 3.393 & 0.369 & 0.258 & 0.138 & 0.180 & 0.167 & 0.254 & 0.449 & 3.151 & 0.378 & 0.455 & 4.197 & 0.374 & 0.097 & 0.410 & 0.156 \\
 & Llama 3.3 & 0.495 & 4.727 & 0.351 & 0.258 & 0.142 & 0.184 & 0.162 & 0.334 & 0.344 & 4.112 & 0.352 & 0.473 & 5.962 & 0.367 & 0.111 & 0.417 & 0.175 \\
 & Qwen 3 & 0.496 & 3.675 & 0.325 & 0.269 & 0.134 & 0.179 & 0.149 & 0.297 & 0.438 & 2.957 & 0.364 & 0.475 & 5.050 & 0.360 & 0.106 & 0.396 & 0.167 \\
 & Gemma 4 & 0.493 & 3.865 & 0.352 & 0.270 & 0.190 & 0.223 & 0.158 & 0.273 & 0.420 & 2.812 & 0.513 & 0.459 & 4.034 & 0.322 & 0.099 & 0.478 & 0.165 \\
 & Gemini 3.5 & 0.492 & 2.978 & 0.368 & 0.263 & 0.149 & 0.190 & 0.181 & 0.303 & 0.587 & 1.293 & 0.567 & 0.446 & 3.062 & 0.325 & 0.092 & 0.437 & 0.152 \\
 & GPT 5.6 & 0.492 & 4.122 & 0.348 & 0.290 & 0.176 & 0.219 & 0.187 & 0.315 & 0.552 & 2.107 & 0.529 & 0.456 & 4.921 & 0.341 & 0.118 & 0.481 & 0.189 \\
\midrule
\multirow{8}{*}{Qwen 3-o} & K2-V2 & 0.489 & 2.103 & 0.108 & 0.106 & 0.091 & 0.098 & 0.054 & 0.178 & 0.290 & 1.107 & 0.123 & 0.330 & 1.507 & 0.106 & 0.037 & 0.317 & 0.066 \\
 & OLMo 3.1 & 0.521 & 5.275 & 0.230 & 0.269 & 0.177 & 0.214 & 0.147 & 0.350 & 0.339 & 3.980 & 0.275 & 0.501 & 6.604 & 0.216 & 0.120 & 0.501 & 0.194 \\
 & Qwen 2.5 & 0.502 & 4.486 & 0.364 & 0.315 & 0.120 & 0.173 & 0.200 & 0.252 & 0.367 & 5.021 & 0.362 & 0.477 & 4.863 & 0.364 & 0.124 & 0.410 & 0.190 \\
 & Llama 3.3 & 0.490 & 5.161 & 0.373 & 0.301 & 0.145 & 0.196 & 0.152 & 0.324 & 0.346 & 4.721 & 0.351 & 0.480 & 6.173 & 0.371 & 0.131 & 0.436 & 0.202 \\
 & Qwen 3 & 0.504 & 4.564 & 0.334 & 0.317 & 0.135 & 0.189 & 0.189 & 0.306 & 0.357 & 4.432 & 0.345 & 0.486 & 5.786 & 0.357 & 0.133 & 0.427 & 0.203 \\
 & Gemma 4 & 0.511 & 4.845 & 0.370 & 0.315 & 0.166 & 0.217 & 0.183 & 0.271 & 0.360 & 4.593 & 0.442 & 0.471 & 4.713 & 0.317 & 0.116 & 0.448 & 0.185 \\
 & Gemini 3.5 & 0.486 & 3.975 & 0.362 & 0.292 & 0.151 & 0.199 & 0.163 & 0.297 & 0.394 & 2.824 & 0.498 & 0.452 & 3.883 & 0.308 & 0.103 & 0.434 & 0.167 \\
 & GPT 5.6 & 0.512 & 5.143 & 0.363 & 0.347 & 0.139 & 0.198 & 0.196 & 0.278 & 0.407 & 4.236 & 0.434 & 0.483 & 5.484 & 0.319 & 0.143 & 0.438 & 0.216 \\
\midrule
\multirow{8}{*}{Qwen 3-o Cap.} & K2-V2 & 0.085 & 1.052 & 0.007 & 0.008 & 0.077 & 0.015 & 0.000 & 0.214 & 0.031 & 0.808 & 0.010 & 0.057 & 0.995 & 0.007 & 0.002 & 0.353 & 0.004 \\
 & OLMo 3.1 & 0.513 & 4.923 & 0.177 & 0.263 & 0.187 & 0.218 & 0.125 & 0.369 & 0.274 & 3.551 & 0.135 & 0.506 & 5.325 & 0.131 & 0.113 & 0.488 & 0.183 \\
 & Qwen 2.5 & 0.494 & 4.718 & 0.360 & 0.322 & 0.115 & 0.169 & 0.151 & 0.263 & 0.345 & 4.615 & 0.258 & 0.492 & 5.414 & 0.390 & 0.127 & 0.372 & 0.189 \\
 & Llama 3.3 & 0.495 & 5.407 & 0.354 & 0.294 & 0.151 & 0.200 & 0.160 & 0.332 & 0.279 & 6.314 & 0.240 & 0.522 & 5.082 & 0.367 & 0.124 & 0.479 & 0.197 \\
 & Qwen 3 & 0.486 & 4.573 & 0.341 & 0.296 & 0.130 & 0.181 & 0.165 & 0.342 & 0.291 & 5.355 & 0.238 & 0.477 & 5.472 & 0.370 & 0.126 & 0.434 & 0.196 \\
 & Gemma 4 & 0.507 & 5.141 & 0.343 & 0.327 & 0.160 & 0.215 & 0.178 & 0.264 & 0.307 & 6.117 & 0.311 & 0.489 & 5.929 & 0.311 & 0.124 & 0.443 & 0.194 \\
 & Gemini 3.5 & 0.503 & 5.023 & 0.329 & 0.308 & 0.133 & 0.186 & 0.192 & 0.270 & 0.389 & 3.622 & 0.440 & 0.495 & 5.785 & 0.305 & 0.119 & 0.420 & 0.185 \\
 & GPT 5.6 & 0.515 & 5.158 & 0.349 & 0.352 & 0.139 & 0.199 & 0.181 & 0.294 & 0.399 & 4.110 & 0.408 & 0.500 & 6.411 & 0.335 & 0.155 & 0.446 & 0.231 \\
\midrule
\multirow{8}{*}{AF 3} & K2-V2 & 0.450 & 2.066 & 0.180 & 0.142 & 0.129 & 0.135 & 0.102 & 0.239 & 0.304 & 1.554 & 0.308 & 0.445 & 1.597 & 0.182 & 0.053 & 0.476 & 0.096 \\
 & OLMo 3.1 & 0.481 & 3.044 & 0.196 & 0.212 & 0.151 & 0.176 & 0.136 & 0.340 & 0.360 & 3.327 & 0.284 & 0.469 & 4.346 & 0.185 & 0.094 & 0.492 & 0.158 \\
 & Qwen 2.5 & 0.484 & 2.614 & 0.378 & 0.256 & 0.155 & 0.193 & 0.187 & 0.297 & 0.465 & 2.673 & 0.436 & 0.420 & 2.848 & 0.309 & 0.086 & 0.508 & 0.148 \\
 & Llama 3.3 & 0.502 & 3.877 & 0.388 & 0.272 & 0.160 & 0.201 & 0.185 & 0.368 & 0.352 & 4.382 & 0.458 & 0.452 & 4.317 & 0.353 & 0.102 & 0.496 & 0.170 \\
 & Qwen 3 & 0.484 & 2.667 & 0.322 & 0.265 & 0.156 & 0.196 & 0.160 & 0.330 & 0.464 & 2.483 & 0.402 & 0.421 & 3.026 & 0.351 & 0.098 & 0.469 & 0.162 \\
 & Gemma 4 & 0.508 & 2.608 & 0.398 & 0.262 & 0.218 & 0.238 & 0.160 & 0.304 & 0.547 & 1.588 & 0.512 & 0.430 & 2.822 & 0.311 & 0.085 & 0.560 & 0.147 \\
 & Gemini 3.5 & 0.493 & 2.319 & 0.397 & 0.276 & 0.175 & 0.215 & 0.198 & 0.340 & 0.612 & 1.133 & 0.571 & 0.429 & 2.686 & 0.325 & 0.087 & 0.512 & 0.149 \\
 & GPT 5.6 & 0.500 & 2.863 & 0.383 & 0.273 & 0.182 & 0.218 & 0.191 & 0.340 & 0.640 & 1.284 & 0.527 & 0.428 & 3.429 & 0.327 & 0.096 & 0.510 & 0.162 \\
\bottomrule
\end{tabular}
}
\end{table*}
\Cref{tbl:casc} reports all 48 cascaded systems, \textit{i.e.}, six captioners paired with eight text models, on Ego4D, SINS, and EgoLife.
\Cref{tbl:audiocap} and \Cref{tbl:llm} average this grid along each axis.

\subsection{Class-wise frame accuracy of top-line systems}\label{apdx:facc}
\begin{table*}[t!]
\centering
\caption{Comparison of frame-wise accuracy (fAcc) for each event class across top-line systems. Cascaded systems use Audio Flamingo 3 as the audio captioning model. The Ratio column indicates the proportion of frames belonging to each class within the dataset. Classes are sorted by ratio.}
\label{tbl:facc}

\resizebox{\textwidth}{!}{
\begin{tabular}{l c| c c c c| c c c c}
\toprule
\textbf{Ego4D} & & \multicolumn{4}{c|}{\textbf{Cascaded}} & \multicolumn{4}{c}{\textbf{End-to-end}} \\
\textbf{Class} & \textbf{Ratio} & \textbf{Gemini 3.5} & \textbf{GPT 5.6} & \textbf{Gemma 4} & \textbf{Gemma 4 + SFT} & \textbf{Gemini 3.5} & \textbf{Qwen 3-o} & \textbf{Qwen 3-o + SFT} & \textbf{Qwen 2.5-o} \\
\midrule
productive & 0.401 & 0.369 & 0.336 & 0.324 & 0.603 & 0.503 & 0.269 & 0.763 & 0.945 \\
leisure    & 0.285 & 0.601 & 0.616 & 0.688 & 0.695 & 0.513 & 0.841 & 0.680 & 0.017 \\
food       & 0.117 & 0.460 & 0.408 & 0.451 & 0.532 & 0.447 & 0.229 & 0.646 & 0.000 \\
traveling  & 0.082 & 0.157 & 0.164 & 0.138 & 0.054 & 0.114 & 0.042 & 0.211 & 0.132 \\
other      & 0.059 & 0.086 & 0.133 & 0.055 & 0.000 & 0.191 & 0.264 & 0.000 & 0.000 \\
purchasing & 0.056 & 0.098 & 0.072 & 0.088 & 0.232 & 0.434 & 0.348 & 0.720 & 0.067 \\
\bottomrule
\end{tabular}
}

\vspace{1em}
\resizebox{\textwidth}{!}{
\begin{tabular}{l c| c c c c| c c c c}
\toprule
\textbf{EgoLife} & & \multicolumn{4}{c|}{\textbf{Cascaded}} & \multicolumn{4}{c}{\textbf{End-to-end}} \\
\textbf{Class} & \textbf{Ratio} & \textbf{Gemini 3.5} & \textbf{GPT 5.6} & \textbf{Gemma 4} & \textbf{Gemma 4 + SFT} & \textbf{Gemini 3.5} & \textbf{Qwen 3-o} & \textbf{Qwen 3-o + SFT} & \textbf{Qwen 2.5-o} \\
\midrule
productive & 0.643 & 0.173 & 0.145 & 0.129 & 0.127 & 0.262 & 0.013 & 0.136 & 0.970 \\
leisure    & 0.268 & 0.712 & 0.778 & 0.777 & 0.849 & 0.418 & 0.857 & 0.671 & 0.013 \\
purchasing & 0.039 & 0.025 & 0.015 & 0.020 & 0.200 & 0.652 & 0.544 & 0.805 & 0.018 \\
traveling  & 0.029 & 0.272 & 0.272 & 0.226 & 0.151 & 0.491 & 0.221 & 0.705 & 0.304 \\
food       & 0.019 & 0.767 & 0.719 & 0.691 & 0.764 & 0.602 & 0.386 & 0.737 & 0.059 \\
other      & 0.003 & 0.117 & 0.186 & 0.052 & 0.000 & 0.188 & 0.208 & 0.000 & 0.000 \\
\bottomrule
\end{tabular}
}

\vspace{1em}
\resizebox{\textwidth}{!}{
\begin{tabular}{l c| c c c c| c c c c}
\toprule
\textbf{SINS} & & \multicolumn{4}{c|}{\textbf{Cascaded}} & \multicolumn{4}{c}{\textbf{End-to-end}} \\
\textbf{Class} & \textbf{Ratio} & \textbf{Gemini 3.5} & \textbf{GPT 5.6} & \textbf{Gemma 4} & \textbf{Gemma 4 + SFT} & \textbf{Gemini 3.5} & \textbf{Qwen 3-o} & \textbf{Qwen 3-o + SFT} & \textbf{Qwen 2.5-o} \\
\midrule
sleeping      & 0.273 & 0.037 & 0.023 & 0.102 & 0.057 & 0.349 & 0.088 & 0.181 & 0.016 \\
watching tv  & 0.226 & 0.663 & 0.511 & 0.391 & 0.762 & 0.774 & 0.722 & 0.948 & 0.760 \\
absence       & 0.171 & 0.975 & 0.978 & 0.912 & 0.000 & 0.483 & 0.914 & 0.000 & 0.402 \\
working       & 0.171 & 0.902 & 0.886 & 0.894 & 0.986 & 0.350 & 0.189 & 0.980 & 0.034 \\
cooking       & 0.035 & 0.571 & 0.410 & 0.468 & 0.727 & 0.615 & 0.602 & 0.587 & 0.139 \\
other         & 0.025 & 0.502 & 0.528 & 0.504 & 0.352 & 0.399 & 0.434 & 0.070 & 0.553 \\
calling       & 0.020 & 0.602 & 0.794 & 0.671 & 0.124 & 0.613 & 0.270 & 0.190 & 0.033 \\
eating        & 0.017 & 0.601 & 0.599 & 0.501 & 0.240 & 0.769 & 0.170 & 0.384 & 0.058 \\
visit         & 0.013 & 0.786 & 0.723 & 0.806 & 0.276 & 0.250 & 0.179 & 0.735 & 0.000 \\
toilet        & 0.011 & 0.197 & 0.168 & 0.177 & 0.000 & 0.040 & 0.000 & 0.000 & 0.010 \\
dishwashing   & 0.011 & 0.263 & 0.286 & 0.312 & 0.379 & 0.512 & 0.138 & 0.000 & 0.000 \\
vacuumcleaner & 0.007 & 0.909 & 0.867 & 0.938 & 0.942 & 0.747 & 0.081 & 0.000 & 0.000 \\
showering     & 0.007 & 0.778 & 0.729 & 0.843 & 0.785 & 0.713 & 0.348 & 0.879 & 0.318 \\
dressing      & 0.005 & 0.138 & 0.078 & 0.063 & 0.139 & 0.416 & 0.136 & 0.094 & 0.003 \\
toothbrushing & 0.003 & 0.893 & 0.825 & 0.895 & 0.248 & 0.621 & 0.000 & 0.000 & 0.441 \\
shaving       & 0.003 & 0.934 & 0.944 & 0.933 & 0.850 & 0.694 & 0.582 & 0.870 & 0.000 \\
getting dry  & 0.002 & 0.710 & 0.577 & 0.642 & 0.091 & 0.070 & 0.092 & 0.000 & 0.000 \\
\bottomrule
\end{tabular}
}

\end{table*}

\Cref{tbl:facc} reports the class-wise fAcc for each class, ordered in frequency, \textit{i.e.}, ratio of the event on the full audio.
Class-wise fAcc is the fraction of frames of that class the system labels correctly.

\end{document}